\documentclass[12pt]{article}

\usepackage{amssymb}
\usepackage{amsmath}

\usepackage{cite}

\begin{document}

\title{\bf On Superplanckian Scattering on the Brane}
\author{A.V. Kisselev\thanks{Email address: kisselev@th1.ihep.su} \\
\small \em  Institute for High Energy Physics, 142281 Protvino,
Russia}

\date{}

\maketitle

\begin{abstract}
The multidimensional space-time with $(D-4)$ compact extra space
dimensions and SM fields confined on four-dimensional brane is
considered. The elastic scattering amplitude of two particles
interacting by gravitational forces is calculated at
superplanckian energies. A particular attention is paid to a
proper account of zero (massless) graviton mode. The renormalized
Born pole is reproduced in the eikonal amplitude which makes a
leading contribution at small momentum transfers. This singular
part of the amplitude coincides with well-known $D$-dimensional
amplitude taken at $D \rightarrow 4$. The expression for a
contribution from massive graviton modes to the eikonal is
derived, and it asymptotics in the impact parameter are
calculated. Our formula gives correct four-dimensional result at
$R_c \rightarrow 0$, where $R_c$ is the radius of the higher
dimensions, contrary to formulae obtained in recent papers on
collisions of particles living on the brane. The results are also
compared with those obtained previously for the scattering of the
bulk fields in flat higher dimensions.
\end{abstract}

\section{Introduction}

In the 4-dimensional space-time the gravity is very weak as
compared with the interactions of the Standard Model (SM) fields.
Namely, the Newton constant is equal to $G_N = M_{Pl}^{-2}$, where
$M_{Pl} = 1.2 \cdot 10^{19}$ GeV is the Planck mass, while the
electroweak scale is about $m_{EW} \sim 10^3$ GeV. In order to
explain a huge ratio of two physical scales in nature,
$M_{Pl}/m_{EW}$, a scheme with additional space dimensions with a
flat metric has been proposed~\cite{Arkani-Hamed:98} (in what
follows, referred to as a ADD model). All $d$ extra dimensions are
compact with the radius $R_c$. In other words, the space-time is
$R^4 \times M_d$, where $M_d$ is a $d$-dimensional manifold of the
volume $R_c^d$. If $R_c^{-1} \ll m_{EW}$, a gravitational
potential will get negligible corrections at distances $r \gg
R_c$.

Let $M_D$ be a fundamental Planck scale in $D$-dimensional theory
($D=4+d$). Then it can be shown~\cite{Arkani-Hamed:98} that
\begin{equation}
M_{Pl}^2 = R_c^d M_D^{2+d}.
\label{10}
\end{equation}
One can get $M_D \sim 1$ TeV, if the compactification radius $R_c$
is large enough. The radius $R_c$ depends on $d$ and it ranges
from $1$ mm to $1$ fm if $d$ runs from $2$ to $6$. Since $R_c \gg
m_{EW}^{-1}$, all Standard Model (SM) gauge and matter fields are
to be confined to a $3$-dimensional brane embedded into the
$(3+d)$-dimensional space (gravity alone lives in the bulk).

From the point of view of the 4-dimensional space-time, there
arise a Kaluza-Klein (KK) tower of massive graviton modes, $G_{\mu
\nu}^{(n)}$, with masses
\begin{equation}
m_n = \frac{\sqrt{n^2}}{R_c} \, , \quad n^2 = n_1^2 + n_2^2 +
\ldots + n_d^2,
\label{12}
\end{equation}
where $n$ defines the KK excitation level. So, a mass splitting is
$\Delta m \sim R_c^{-1}$ and we have an almost continuous spectrum
of the gravitons.

The interaction of the gravitons with the SM fields is described
by the Lagrangian~\cite{Arkani-Hamed:98}
\begin{equation}
\mathcal{L} = -\frac{1}{{\bar M}_{Pl}} G_{\mu \nu}^{(n)} T^{\mu
\nu},
\label{13}
\end{equation}
where $\mu,\nu = 0,1,2,3$ and $\bar{M}_{Pl} = M_{Pl}/\sqrt{8\pi}$
is the reduced Planck mass. One can conclude from (\ref{13}) that
the coupling of both massless and massive graviton is universal
and very small ($\sim 1/{\bar M}_{Pl}$). However, the multiplicity
of the KK states produced in high energy collisions is huge and it
is equal to $(\sqrt{s} R_c)^d$, where $\sqrt{s}$ is the collision
energy. A typical cross-section of a process involving the
production of the KK graviton excitations with masses $m_n \leq
\sqrt{s}$ is suppressed only by the scale $M_D$:
\begin{equation}
\sigma_{KK} \sim \frac{s^{d/2}}{M_D^{d+2}}.
\label{14}
\end{equation}
So, the ADD model can be tested at future hadronic colliders and
at $e^+e^-$ linear colliders in the range of TeV energies (the
Planck regime). There is a lot of papers on a collider
phenomenology within the framework of the extra dimensions. An
interested reader can find references in
reviews~\cite{Kisselev:02}.

The collisions in the transplanckian regime ($\sqrt{s} \gg M_D$)
were considered in a variety of papers in the framework of the
string theory~\cite{Amati:87}, in the eikonal approximation of the
reggeized graviton exchange~\cite{Muzinich:88} as well as in
different 4-dimensional approaches~\cite{Hooft:87,Verlinde:92}.
The equivalence of various schemes has been demonstrated
in~\cite{Kabat:92}. Note, in papers~\cite{Amati:87,Muzinich:88} a
collision of the \emph{bulk fields} in the $D$-dimensional
\emph{flat space-time} was considered.

In Refs.~\cite{Nussinov:99} an estimate of high-energy
gravitational cross sections of hadrons have been made. The
contribution from the KK excitations of the graviton changes
$t$-channel propagator $(-t)^{-1}$ by
\begin{equation}
\frac{1}{-t} \; \rightarrow \sum_{n_1^2 + \ldots n_d^2 \geqslant
0} \frac{1}{-t + \sum\limits_{i=1}^d \displaystyle
\frac{n_i^2}{\displaystyle R_c^2}}.
\label{I5}
\end{equation}
It was argued in \cite{Nussinov:99}, that the range $n \lesssim
n_{max} \sim M_s R_c$, where $M_s$ is a quantum gravity (string)
scale, makes a dominant contribution to hadronic cross sections.
Using a replacement ($d > 2$)
\begin{equation}
\sum_{n_1^2 + \ldots n_d^2 \geqslant 0} \frac{1}{-t +
\sum\limits_{i=1}^d \displaystyle \frac{n_i^2}{\displaystyle
R_c^2}} \; \rightarrow \; R_c^2 \int \! d^d \Omega
\int\limits_0^{n_{max}} \! dn \, n^{d-3},
\label{16}
\end{equation}
the following results for the total cross section has been
obtained (after unitarization):
\begin{equation}
\sigma_{tot}(s) \simeq \frac{4\pi s}{M_D^4}.
\label{17}
\end{equation}

However, papers~\cite{Nussinov:99} have unjustified approximations
and incorrect estimates, as it was pointed out in
Ref.~\cite{Petrov:02}. In particular, the presence of the massless
exchange quantum (zero mode of the graviton) should result in
\emph{infinite elastic and total hadronic cross sections},
contrary to equality (\ref{17}). For strong interactions without
gravitational forces, the upper (Froissart) bound for
$\sigma_{tot}(s)$ is modified by an additive term $(\pi
r_c/m_{\pi}) \ln s$, if the extra dimensions are compactified onto
a circle with the radius $r_c$~\cite{Petrov:01}.

Recently, results on a collision of the \emph{brane particles},
which interact by graviton forces in the ADD model with the
compact extra dimensions, have been presented in
\cite{Giudice:02}.

We are confident, however, that the massless graviton mode was not
taken properly into account in both \cite{Nussinov:99} and
\cite{Giudice:02}. In particular, we can not agree with a
conclusion that long-range forces are completely hidden by the
interactions of the massive graviton excitations, coming from the
extra dimensions \cite{Giudice:02}. That is why, in the present
paper we calculate the scattering amplitude for two particles
confined on the brane, by separating massless graviton
contribution from massive graviton effects from the very begining.

In the next Section we review briefly the results on the
scattering in both $D$ flat dimensions~\cite{Amati:87,Muzinich:88}
and four flat dimensions~\cite{Hooft:87,Verlinde:92}. In the
beginning of Section \ref{sec:brane} we consider the approach
proposed in \cite{Giudice:02}. The rest of the
Section~\ref{sec:brane} is devoted to calculations of the eikonal
amplitude. In the last Section we discuss our results and compare
them with the results obtained by other authors mentioned in the
paper. In the Appendix technical details of our calculations are
presented.

\section{Transplanckian collision in the bulk}
\label{sec:bulk}

In this Section we remind some results on transplanckian
collisions in models with the extra dimensions. As was mentioned
in the Introduction, the transplanckian regime has been analyzed
in details in the string theory. The string theory has the
fundamental classical constant $\alpha'$, its inverse being the
string tension. Since the leading graviton trajectory is at
$\alpha(t) = 2 + (\alpha'/2)t$, one expects that at high $s$
graviton exchange will dominate light-string scattering amplitude
for any number of loops.

The transplanckian regime is characterized by a strong effective
coupling $\alpha_G(s) = G_D s  $ ($G_D = M_D^{-(2+d)}$ is a
$D$-dimensional Newton constant). In Refs.~\cite{Amati:87}
four-string scattering amplitude was calculated in the kinematical
region
\begin{eqnarray}
\alpha' s &\gtrsim& (M_D \sqrt{\alpha})^{d+2} \gg 1,
\nonumber \\
|t| &\gtrsim& \alpha'^{-1}, \quad \alpha'|t| \gtrsim
(\alpha'R_c)^{-2}.
\label{18}
\end{eqnarray}
Inequalities (\ref{18}) mean that the tree amplitude is large
($G_D s \alpha'^{- d/2} \gg 1$), while the loop expansion
parameter, $G_D \alpha'^{-(1+d/2)}$, is small. Due to the second
restriction on $t$ in (\ref{18}), \emph{compactified momenta are
not noticeably excited}. In terms of an impact parameter $b$, the
limitations look like
\begin{equation}
b > \lambda_s, \quad b > R_G(s),
\label{20}
\end{equation}
where $\lambda_s = \sqrt{2 \alpha' \hbar}$ is a fundamental
quantum length in the string theory and $R_G(s) \simeq (2G_D
\sqrt{s})^{1/(d+1)}$ is a gravitational radius.

The leading contribution to the scattering amplitude at the impact
parameter $b$ has all powers of $\alpha_G(s)$ and it is the same
in all approaches at $b \gg \lambda_{Pl},\,R_G(s)$, where
$\lambda_{Pl} = (\hbar G_D)^{1/(d+2)}$ is the Planck length.

The amplitude is of a classical (eikonal) form. For large $b$ the
eikonal function is given by~\cite{Amati:87}
\begin{equation}
\chi (b,s) \equiv \chi_{_{ACV}}(b,s) \simeq \left(
\frac{\tilde{b}_c}{b} \right)^d + i \pi^2 \frac{G_D s
\alpha'^{-d/2}}{(\pi \ln s)^{d/2+1}} \exp \left( - \frac{b^2}{4
\alpha' \ln s} \right),
\label{22}
\end{equation}
where $\tilde{b}_c = [\alpha_G(s) 2 \pi^{-d/2}
\Gamma(d/2)]^{1/d}$. As one can see, \emph{$\chi_{_{ACV}}(b,s)$
has both real and imaginary part}. The former has a power-like
behavior in $b$, while the latter decreases exponentially at $b
\gg 2\alpha' \ln s$. Correspondingly, at small $t$ (namely, at
$|t| \lesssim \alpha_G(s)^{-2/d}$) the amplitude has the
asymptotics
\begin{eqnarray}
A^{eik}_{_{ACV}}(s,t) &\simeq&  A_{B}(s,t)
\nonumber \\
&+& i \,\text{const} \frac{(16\pi  \alpha_G(s) )^2 s}{d(d - 2)}
\left[ - |t|^{d/2-1} +  (16\pi G_D s)^{2/d -1}\right].
\label{24}
\end{eqnarray}
Here
\begin{equation}
A_{B}(s,t) = \frac{8\pi \alpha_G(s) s}{-t}
\label{26}
\end{equation}
is the Born amplitude. Thus, \emph{the Born term dominates at
small $t$}.

For large $t$ ($|t| \gg \tilde{b}_c^{-2}$) the amplitude has the
following behavior~\cite{Amati:87}:
\begin{eqnarray}
A^{eik}_{_{ACV}}(s,t) &\sim& \frac{8\pi \alpha_G(s) s}{-t} \,
\exp(i \phi_D)
\nonumber \\
&\times& \left(  4\pi (\tilde{b_c} \sqrt{|t|})^d
\right)^{-d/2(d+1)}.
\label{28}
\end{eqnarray}
The $D$-dimensional phase
\begin{equation}
\phi_D = \frac{d+1}{d} \left( G_D s \, 2\pi^{-d/2} \Gamma(1 + d/2)
|t|^{d/2} \right)^{1/(d+1)}
\label{30}
\end{equation}
has a pole at $D=4$. So, the limit  $D \rightarrow 4$ is
completely non-perturbative due to the divergent (Coulomb) phase.

In Ref.~\cite{Muzinich:88} the same result (\ref{28}) was obtained
by summing multiple reggeized graviton exchange in the eikonal
approximation. Although Regge behavior is present at each order,
it is absent in the final result (\ref{28}). It is important to
note that the magnitude of the \emph{scattering amplitude is
defined by a single non-reggeized graviton exchange} (in full
analogy with the case of Coulomb scattering dominated by a single
photon exchange).

The scattering amplitude may be directly calculated in four
dimensions~\cite{Hooft:87,Verlinde:92} (see also
\cite{Fabbrichesi:94}):
\begin{equation}
A_{HVV}^{eik}(s,t) = A_{B}(s,t) \frac{\Gamma(1 -
i\alpha_G(s))}{\Gamma(1 + i\alpha_G(s))} \left( \frac{4
\mu_{IR}^2}{-t} \right)^{-i\alpha_G(s)},
\label{32}
\end{equation}
or it can be obtained from the $D$-dimensional expression by
taking the limit $D \rightarrow 4$~\cite{Amati:87,Muzinich:88}.
The quantity $\mu_{IR}$ in (\ref{32}) is an infrared cutoff. It
arises in the limit $D \rightarrow 4$, when the pole $(D -
4)^{-1}$ is interpreted as the logarithm of
$\mu_{IR}$~\cite{Muzinich:88}. If the amplitude is calculated as a
sum of soft gravitons with a small mass $m_{grav}$, this cutoff is
related to a graviton mass, $\mu_{IR} = (1/2)m_{grav} e^{\gamma}$,
where $\gamma$ is the Euler constant.

Let us note, the eikonal amplitude in quantum electrodynamics can
be obtained from (\ref{32}) by a simple replacement $-\alpha_G(s)
\rightarrow \alpha_{em} = e_1e_2/4\pi$, where $e_{1,2}$ are
electric charges of colliding massless charged
particles~\cite{Jackiw:92}. In such a case, $\mu_{IR}$ is
proportional to a regulating photon "mass"~\cite{Jackiw:92}.

At large $z$, $|\arg z| < \pi$, the $\Gamma$-function has an
asymptotics $\Gamma(z) = \sqrt{2\pi} e^{-z} e^{(z - 1/2) \ln z} [1
+ \mbox{O}(z^{-1})]$~\cite{Bateman}. Then we obtain from
(\ref{32}) that in four dimensions (see also \cite{Muzinich:88})
\begin{eqnarray}
A^{eik}(s,t) \Big|_{\alpha_G(s) \gg 1} \Big. &\simeq& -i
A_{B}(s,t) \left( \frac{4 \mu_{IR}^2}{-t} \right)^{-i\alpha_G(s)}
\nonumber \\
&\times& \exp \, [-i 2\alpha_G(s) (\ln \alpha_G(s) - 1)].
\label{34}
\end{eqnarray}

In the next Section we will consider the case when colliding
particles are confined to the brane, with the gravity living in
the bulk. Another difference will be that the compactified momenta
become essential, contrary to the approach considered in this
Section.

\section{Transplanckian collision on the brane}
\label{sec:brane}

In the ADD model, all SM fields live on the $(1 + 3)$-dimensional
brane embedded in the $D$-dimensional space-time. Thus, their
collisions are also confined to the brane. In particular, the
impact parameter space is two-fold. On the other hand, in the
transplanckian region, where the collision energy $\sqrt{s}$ is
much larger than the fundamental gravity scale $M_D$, but a
momentum transfer $t$ is small, the scattering of 4-dimensional
particles is dominated by the exchange of $D$-dimensional
gravitons.

The (elastic) scattering of two (different) massless particles
\emph{living on the brane} in the kinematical region
\begin{equation}
\sqrt{s} \gg M_D, \qquad s \gg -t
\label{36}
\end{equation}
was recently considered in Refs.~\cite{Giudice:02,Giudice*:02}.
The ladder diagrams contributing to a \emph{nonreggeized graviton
exchange} in $t$-channel were summed in the eikonal
approximation~\cite{Giudice:02}. From the point of view of a
4-dimensional observer, the massless bulk graviton is represented
by the tower of massive gravitons (\ref{12}). Since the higher
space dimensions are compactified with the radius $R_c$, one has a
sum in a (quantized) momentum transfer in the extra dimensions
$q_{\bot}^{(n)} = n/R_c$ instead of an integral in $d^{D-4}
q_{\bot}$. The Born amplitude is, therefore, of the form
\begin{equation}
A^B(s,t) = G_N s^2 \!\!\!\!\! \sum_{n_1^2 + \ldots n_d^2 \geqslant
0} \frac{1}{ -t + \sum\limits_{i=1}^d  \displaystyle \frac{n_i^
2}{\displaystyle R_c^2}}.
\label{38}
\end{equation}
Here and in what follows the reduced gravitational constant,
$\bar{G}_N = \bar{M}_{Pl}^{-2}$ is always assumed (\ref{13}). For
simplicity, we will write $G_N $ instead of $\bar{G}_N$ (and,
correspondingly, $G_D$ instead of $\bar{G}_D$). Thus, to compare
our results this those of
Refs.~\cite{Amati:87,Muzinich:88,Hooft:87,Verlinde:92}, one will
have to use a substitution $G_N \rightarrow 8\pi G_N $.

In Ref.~\cite{Giudice:02} the following replacement
\begin{equation}
\sum_{n_1^2 + \ldots n_d^2 \geqslant 0} \frac{1}{-t +
\sum\limits_{i=1}^d \displaystyle \frac{n_i^2}{\displaystyle
R_c^2}} \; \rightarrow \; \int \! d^d \Omega
\int\limits_0^{\infty} \! dl l^{d-1} \frac{1}{-t + \displaystyle
\frac{l^2}{\displaystyle R_c^2}}
\label{40}
\end{equation}
was made by assuming that $R_c$ is large (compare with
(\ref{16})). As a result, it was obtained:
\begin{equation}
A_{_{GRW}}^B(s,t) = \pi^{d/2} \Gamma(1 - d/2) \left(
\frac{s}{M_D^2} \right)^2 \left( \frac{-t}{M_D^2} \right)^{d/2 -
1}.
\label{42}
\end{equation}

At one and higher-loop levels it is ladder diagram that makes a
leading contribution to the amplitude. The sum of all such
diagrams results in the eikonal representation for the
amplitude~\cite{Giudice:02}:
\begin{equation}
A^{eik}_{_{GRW}}(s,t) = -2is \int d^2 b_{\bot}  e^{i q_{\bot}
b_{\bot}} \left (e^{i \chi_{_{GRW}}(b_{\bot})} - 1 \right),
\label{44}
\end{equation}
with the eikonal given by
\begin{equation}
\chi_{_{GRW}}( b_{\bot}) = \frac{1}{2s} \int \frac{d^2
q_{\bot}}{(2\pi)^2} e^{-i q_{\bot} b_{\bot}} A^B(s, q_{\bot}^2).
\label{46}
\end{equation}
After a substitution of the Born amplitude (\ref{42}) in equation
(\ref{46}), one gets~\cite{Giudice:02}
\begin{equation}
\chi_{_{GRW}}(b) = \left( \frac{b_c}{b} \right)^d,
\label{48}
\end{equation}
where
\begin{equation}
b_c = \left[ \frac{s \,(4\pi)^{d/2-1} \Gamma(d/2)}{2 M_D^{d+2}}
\right]^{1/d} \equiv 2 \sqrt{\pi} R_c \left[ \frac{G_N s \,
\Gamma(d/2)}{8\pi} \right]^{1/d}.
\label{50}
\end{equation}
At $d \rightarrow 0$, the eikonal $\chi_{_{GRW}}(b)$ (\ref{48})
has an expansion:
\begin{equation}
\chi_{_{GRW}}(b) \Big|_{d \rightarrow 0} \simeq \frac{G_N s}{8\pi}
\left[ \frac{2}{d} + \ln \bigg( \frac{2R_c}{b} \bigg)^2 + \Psi(1) +
\ln \pi \right],
\label{51}
\end{equation}
where $\Psi(z)$ is the $\Psi$-function~\cite{Bateman}.

At all intermediate steps (\ref{40})-(\ref{46}), the number of the
extra dimension $d$ was regarded as a (non integer) parameter. The
final expression (\ref{48}) has no divergences in $d$ at $d>0$,
although the Born amplitude has simple poles at $d=2, 4 \ldots$
(because of the $\Gamma$-function in (\ref{42})).

In \cite{Giudice*:02} the conclusion was made that ``\emph{even
for $q = 0$, the scattering amplitude is dominated by $b \sim b_c$
and not by $b = \infty$, as opposed to the Coulomb case. This
result follows from the different dimensionalities of the space on
which the scattered  particles and exchange graviton live}''.

We will show that \emph{this statement is not correct} and that
the Born amplitude survives after summation of the KK-excitations
of the graviton and does contribute to the eikonal. In its turn,
this means that long-range forces (Coulomb singularity) still
presents in the scattering of brane particles.

Indeed, for $d = 1$ series (\ref{38}) converges, and it has a pole
$t^{-1}$ corresponding to a zero massless mode of the graviton. It
would be strange to expect that long-range forces do present for
$d=1$, but disappear when the gravity live in more than one extra
dimensions. For $d \geqslant 2$, sum (\ref{38}) is divergent and
it needs a regularization. Following~\cite{Giudice:02}, we will
use the dimensional regularization, by considering $d$ to be
non-integer at intermediate steps of our calculations. The final
result will be well-defined for all $d \geqslant 0$.

Although the change ``\emph{summation in $n$}'' $\rightarrow$
``\emph{integration in $dn$}'' (where $n$ labels the KK excitation
level of the graviton) is justified at $R_c\sqrt{|t|} \gg 1$, it
should be done more accurately than it was dealt with in
(\ref{16}) and (\ref{40}). The crucial point is that \emph{a
contribution from the zero (massless) graviton mode must be
isolated before replacement (\ref{40})}:
\begin{equation}
A^B(s,t) = \frac{G_N s^2 }{-t} +  G_N s^2 \!\!\!\!\! \sum_{n_1^2 +
\ldots n_d^2 \geqslant 1} \frac{1}{-t + m_n^2}.
\label{52}
\end{equation}

In the case of large extra dimensions, when the mass splitting is
small ($\Delta m = 1/R_c$), we can write
\begin{equation}
\sum_{n_1^2 + \ldots n_d^2 \geqslant 1} \frac{1}{-t + m_n^2} \;
\rightarrow \; R_c^d \,\int \! d^d \Omega
\int\limits_{R_c^{-1}}^{\infty} \! dm \,m^{d-1} \frac{1}{-t + m^2}
\label{54}
\end{equation}
(strictly speaking, the inequality  $\sqrt{|t|} R_c \gg 1$ must be
satisfied). Then equation (\ref{52}) can be recast as follows:
\begin{eqnarray}
A^B(s,t) &=& \frac{G_N s^2}{-t} \left[1 + (\sqrt{|t|} R_c)^d
\frac{2\pi^{d/2}}{\Gamma(d/2)} \int\limits_{(\sqrt{|t|}
R_c)^{-1}}^{\infty} \! dy y^{d-1} \frac{1}{1 + y^2} \right].
\nonumber \\
&\equiv& A_{0}^B(s,t) + A_{mass}^B(s,t)
\label{56}
\end{eqnarray}
The integral in the RHS of (\ref{56}), representing \emph{the
contribution from the massive gravitons}, can be calculated and
rewritten in the form
\begin{eqnarray}
&& A_{mass}^B(s,t) = \frac{G_N s^2}{-t} (\sqrt{|t|} R_c)^d
\frac{2\pi^{d/2}}{\Gamma(d/2)} \int\limits_{(\sqrt{|t|}
R_c)^{-1}}^{\infty} \! dy y^{d-1} \frac{1}{1 + y^2}
\nonumber \\
&& = G_N s^2 R_c^2 \frac{\pi^{d/2}}{\Gamma(d/2)(1-d/2)} \ {}_2F_1
\left(1, 1 - \frac{d}{2}; 2 - \frac{d}{2}; tR_c^2 \right),
\label{58}
\end{eqnarray}
where ${}_2F_1 \left( \alpha_1, \alpha_2; \,\beta_1; \, z \right)$
is the hypergeometric function~\cite{Bateman}, and we have taken
into account the relation
\begin{equation}
R_c = \frac{1}{M_D} \left( \frac{M_{Pl}}{M_D} \right)^{2/d} =
\left( \frac{G_D}{G_N} \right)^{1/d}.
\label{60}
\end{equation}
Thus, $A_{mass}^B(s,t)$ converges at $t \rightarrow 0$.

We see from (\ref{58}) that $A_{mass}^B(s,t) = 0$ for $d = 0$. In
such a case, it is the 4-dimensional massless graviton that
contributes to $A^B(s,t)$~(\ref{56}), as it should be.

In order to get an asymptotics of  the Born amplitude at large
$t$, we use the equivalent expression for $A_{mass}^B(s,t)$:
\begin{eqnarray}
A_{mass}^B(s,t) &=&  \frac{G_N s^2}{-t} \pi^{d/2}
\left[ \Gamma(1-d/2) \left( -t R_c^2 \right)^{d/2} \right.
\nonumber \\
&-& \left. \frac{1}{\Gamma(1 + d/2)} \ {}_2F_1 \left(1,
\frac{d}{2}; 1 + \frac{d}{2}; \frac{1}{tR_c^2} \right) \right].
\label{62}
\end{eqnarray}
As it follows from (\ref{62}), large $t$ behavior of
$A_{mass}^B(s,t)$ is similar to that of $A_{_{GRW}}^B(s,t)$
(\ref{42}):
\begin{eqnarray}
A_{mass}^B(s,t) \Big|_{R_c^2|t| \gg 1} &\simeq& \pi^{d/2} \Gamma(1
- d/2) \left( \frac{s}{M_D^2} \right)^2 \left( \frac{-t}{M_D^2}
\right)^{d/2 -
1}
\nonumber \\
&\times& \left[ 1 - \frac{1}{\Gamma(1 + d/2) \Gamma(1 - d/2)} \,\,
(-t R_c^2)^{- d/2} \right].
\label{64}
\end{eqnarray}

It is convenient to divide "massive" part of the eikonal,
\begin{eqnarray}
\chi_{mass}(b) &=& \frac{1}{2s} \int \! \frac{d^2
q_{\bot}}{(2\pi)^2} A_{mass}^B(s,t)
\nonumber \\
&=& \frac{1}{4\pi s} \int\limits_0^{\infty} \! q_{\bot} dq_{\bot}
J_0(b q_{\bot}) A_{mass}^B(s,-q_{\bot}^2),
\label{66}
\end{eqnarray}
into two parts:
\begin{equation}
\chi_{mass} \equiv \chi_{mass}^{(1)}(b) + \chi_{mass}^{(2)}(b).
\label{68}
\end{equation}
Here $\chi_{_{mass}}^{(1)}(b) \equiv \chi_{_{GRW}}(b)$ (\ref{48})
and
\begin{equation}
\chi_{mass}^{(2)}(b) = - G_N s \,\frac{\pi^{d/2-1 }}{8\Gamma(1 +
d/2)} \, I(b),
\label{70}
\end{equation}
where $I(b)$ is given by the integral
\begin{equation}
I(b) = \int\limits_0^{\infty} \! \frac{dx}{x} J_0(\frac{b}{R_c
\sqrt{x}}) \ {}_2F_1 \left(1, \frac{d}{2}; 1 + \frac{d}{2}; - x
\right).
\label{72}
\end{equation}
The integral in (\ref{72}) can not be directly expressed in terms
of algebraic or special functions. But we will be able to
calculate its behavior in impact parameter at both large and small
$b$, if we define $I(b)$ as a limit
\begin{equation}
I = \lim_{\epsilon \rightarrow 0} I_{\epsilon},
\label{74}
\end{equation}
where we have introduced
\begin{equation}
I_{\epsilon}(b) =  \int\limits_0^{\infty} \! dx x^{-1 + \epsilon}
J_0(\frac{b}{R_c \sqrt{x}}) \ {}_2F_1 \left(1, \frac{d}{2}; 1 +
\frac{d}{2}; - x \right).
\label{76}
\end{equation}
The integral in (\ref{76}) is well defined at $-3/4 < \text{Re} \,
\epsilon < 1,\,  \text{Re} \,d/2$ (we assume that $\text{Re}\,d >
0$). Thus, the limit $\lim_{\epsilon \rightarrow 0} I_{\epsilon}$
exists. Moreover, $I_{\epsilon}$ is a table integral (see formula
2.21.4.6 from \cite{Prudnikov}):
\begin{eqnarray}
I_{\epsilon}(b) &=& \frac{1}{\epsilon} \, \frac{\Gamma(1 -
\epsilon)}{\Gamma(1 + \epsilon)} \left[ - \left( \frac{b^2}{4
R_c^2} \right)^{\epsilon} {}_2F_3 \left( \frac{d}{2}, 1; 1 +
\frac{d}{2}, 1 + \epsilon, 1 + \epsilon; \frac{b^2}{4 R_c^2}
\right) \right.
\nonumber \\
&+& \frac{d}{d - 2\epsilon} \, \Gamma^2(1 + \epsilon) \left.
{}_1F_2 \left( \frac{d}{2} - \epsilon; 1 + \frac{d}{2} - \epsilon,
1; \frac{b^2}{4 R_c^2} \right) \right],
\label{78}
\end{eqnarray}
where ${}_pF_q \left( \alpha_1, \ldots \alpha_p; \,\beta_1, \ldots
\beta_q; \, z \right)$ is the generalized hypergeometric
function~\cite{Bateman}.

At $b \ll R_c$, we immediately get from (\ref{74}) and (\ref{78})
($d > 0$):
\begin{eqnarray}
I(b) \Big|_{b \ll R_c} &\simeq& \, \lim_{\epsilon \rightarrow 0}
\frac{1}{\epsilon} \left\{ - \left( \frac{b^2}{4 R_c^2}
\right)^{\epsilon} \left[ 1 + \frac{1}{1 + \epsilon} \,
\frac{d}{d+2} \left( \frac{b}{2R_c} \right)^2  \right]  \right.
\nonumber \\
&+& \left. \frac{d}{d - 2\epsilon} \,   \Gamma^2(1 + \epsilon) \,
\right\}
\nonumber \\
&=& 2 \left\{ \ln \left( \frac{2 R_c}{b} \right) \Big[ 1 +
\frac{d}{d + 2} \left( \frac{b}{2R_c} \right)^2 \Big] +
\frac{1}{d} + \Psi(1) \right\}.
\label{80}
\end{eqnarray}

The region $b \gg R_c$ is much more difficult to analyze. The
asymptotics of $I(b)$ is calculated in the Appendix and the result
is the following:
\begin{eqnarray}
I(b) \Big|_{b \gg R_c} \!\! &\simeq& \, \left( \frac{2 R_c}{b}
\right)^d \frac{\Gamma(1 + d/2)}{\Gamma(1 - d/2)}
\nonumber \\
&\times& \lim_{\epsilon \rightarrow 0} \frac{1}{\epsilon} \left[ -
\frac{\Gamma(1 - d/2)}{\Gamma(1 - d/2 + \epsilon)} +
\frac{\Gamma(d/2 - \epsilon)}{\Gamma(d/2)} \right]
\nonumber \\
&=& \left( \frac{2R_c}{b} \right)^d \Gamma(\frac{d}{2}) \Gamma(1 +
\frac{d}{2}) \cos \frac{\pi d}{2}.
\label{82}
\end{eqnarray}

From formulae (\ref{56}), (\ref{68}) and (\ref{70}) it follows
that
\begin{equation}
\chi(b) = \chi_0(b) + \chi_{mass}(b),
\label{84}
\end{equation}
where
\begin{equation}
\chi_0(b) = \frac{1}{4\pi s} \int\limits_0^{\infty} \!q_{\bot}
dq_{\bot} J_0(b q_{\bot}) A_0^B(s,-q_{\bot}^2)
\label{86}
\end{equation}
and
\begin{equation}
\chi_{mass}(b) = \left( \frac{b_c}{b} \right)^d - G_N s
\,\frac{\pi^{d/2-1 }}{8\Gamma(1 + d/2)} \, I(b).
\label{88}
\end{equation}
The asymptotics of $I(b)$ at small and large $b$ are calculated
above (see (\ref{80}), (\ref{82})). The zero mode graviton
contribution to the Born amplitude is
\begin{equation}
A_0^B(s,t) = \frac{G_N s^2}{-t}.
\label{90}
\end{equation}
Correspondingly, the eikonal amplitude is represented by the
expression
\begin{equation}
A^{eik}(s,t) = - 4\pi i s \int\limits_0^{\infty} \! b \,db J_0(b
\sqrt{-t}) \left[ e^{i(\chi_0(b) + \chi_{mass}(b))} - 1 \right].
\label{96}
\end{equation}

The massless graviton contribution to the eikonal (\ref{86}) is
divergent due to the Coulomb-like pole in $t$ (\ref{90}). In order
to regularize it, let us assume that \emph{the colliding particles
are confined to a $(4+ \delta)$-dimensional brane}  with $\delta
> 0$, while the gravity propagates in $(4 + \delta + d)$
dimensions. Then instead of equation (\ref{86}) we will have
\begin{equation}
\chi_0(b,\delta) = \frac{1}{2 s} \int \! \frac{d^{2+\delta}
q_{\bot}}{(2\pi)^{2+\delta}} \,e^{i b q_{\bot}}
A_0^B(s,-q_{\bot}^2,\delta),
\label{98}
\end{equation}
where $A_0^B(s,t,\delta) = G_{4+\delta}s^2/|t|$ and $G_{4+\delta}$
is a gravitational constant in $(4 + \delta)$ dimensions. The
``massive'' part of the eikonal, $\chi_{mass}(b,\delta)$, is
analogously determined via $A_{mass}(s,t,\delta)$.

We define the (eikonal) amplitude of the scattering in four
dimensions as a limit
\begin{equation}
A^{eik}(s,t) = \lim_{\delta \rightarrow 0} A^{eik}(s,t,\delta),
\label{102}
\end{equation}
where the eikonal amplitude,
\begin{equation}
A^{eik}(s,t,\delta) = - 2i s \int \! d^{2+\delta} b \, e^{i b
q_{\bot}} \left[ e^{i (\chi_0(b,\delta) + \chi_{mass}(b,\delta))}
- 1 \right], \label{104}
\end{equation}
can be rewritten by adding and subtracting the ``massless'' term:
\begin{eqnarray}
A^{eik}(s,t,\delta) &=& - 2i s \int \! d^{2+\delta} b \, e^{i b
q_{\bot}} \, e^{i\chi_0(b,\delta)} \left[
e^{i\chi_{mass}(b,\delta)} - 1 \right]
\nonumber \\
&-& 2i s \int \! d^{2+\delta} b \, e^{i b q_{\bot}} \left[
e^{i\chi_0(b,\delta)} - 1 \right]
\label{106}
\end{eqnarray}

It is easy to check that $\chi_{mass}(t,\delta)$ is non-singular
at $\delta = 0$. As for $\chi_0(t,\delta)$, at small $\delta$ we
get from (\ref{98})
\begin{equation}
\chi_0(b,\delta) \Big|_{\delta \rightarrow 0} = \frac{G_N s}{4\pi}
\left[ \frac{1}{\delta} - \ln (b M_{Pl}) + \text{O}(\delta)
\right].
\label{108}
\end{equation}
The  fisrt  integral in the RHS of (\ref{106}) converges at $b =
0$ ($\chi_{mass}(b) \sim Ab^{-d} + B\ln (1/b)$, if $b \rightarrow
0$), and it is well-defined at $b = \infty$, if $d > 2$
($\chi_{mass}(b) \sim Cb^{-d}$, if $b \rightarrow \infty$).

The second integral in the RHS of (\ref{106}) is well-known. Let
us put
\begin{equation}
\frac{1}{\delta} = \ln \left( \frac{M_{Pl}}{\mu_{IR}} \right),
\label{110}
\end{equation}
where $\mu_{IR}$ is an infrared regulator at $\delta \rightarrow
0$. Then this integral is given by the expression for
$A^{eik}_{HVV}(s,t)$ (\ref{32}) with a replacement $\alpha_G
\rightarrow \alpha_G/8\pi$ (or, equivalently, $G_N \rightarrow
G_N/8\pi$, see our remarks after formula (\ref{38})). As a result,
we obtain
\begin{eqnarray}
&& A^{eik}(s,t)  = \left( \frac{4 \mu_{IR}^2}{-t}
\right)^{-i\alpha_G(s)/8\pi} \Bigg\{ A^B_0(s,t) \, \frac{\Gamma(1
- i\alpha_G(s)/8\pi)}{\Gamma(1 + i\alpha_G(s)/8\pi))}
\nonumber \\
&&  - 4\pi i \frac{s}{-t}  \int\limits_0^{\infty} \, dx\, x \Big(
\frac{x}{2} \Big)^{-i\alpha_G(s)/4\pi} J_0(x) \Big[ \exp \Big( i
\chi_{mass} \Big(\frac{x}{\sqrt{|t|}} \Big) \Big) - 1 \Big]
\Bigg\}. \label{112}
\end{eqnarray}
Here $\chi_{mass}(b)$ is defined by formula (\ref{88}) and
$A^B_0(s,t)$ is the singular part of the Born amplitude
(\ref{90}).

It follows directly from (\ref{66}) and (\ref{58}) that
\begin{equation}
\chi_{mass}(b)\Big|_{d \rightarrow 0} \rightarrow 0.
\label{114}
\end{equation}
As can be seen from (\ref{51}), (\ref{70}) and (\ref{80}), the
small $b$ behavior of $\chi_{mass}(b)$ is consistent with
(\ref{114}). Large $b$ asymptotics of $\chi_{mass}(b)$ also obeys
this limit (see equation (\ref{122})). Thus, by taking the limit
$d \rightarrow 0$ in (\ref{112}), we reproduce the well-known
four-dimensional result (\ref{32}) derived in
Refs.~\cite{Hooft:87,Verlinde:92}.

Introducing a 4-dimensional phase
\begin{equation}
\phi_4 = \frac{G_N s}{8\pi} \ln \left( \frac{-t}{4 \mu_{IR}^2}
\right),
\label{116}
\end{equation}
we get \emph{our final result}:
\begin{eqnarray}
A^{eik}(s,t) &=& s \, e^{i \phi_4} \Big\{ \frac{G_N s}{-t} \,
\frac{\Gamma(1 - i G_N s/8\pi)}{\Gamma(1 + i G_N s/8\pi))} - 16
\pi i \,R_c^2 (R_c \sqrt{|t|})^{-i G_N s/4\pi}
\nonumber \\
&\times& \int\limits_0^{\infty} \, dz \, z^{1 - i G_N s/4\pi}
J_0(2R_c \sqrt{|t|}\,z) \Big[ e^{i \chi_{mass}(z)} - 1 \Big]
\Big\}, \label{118}
\end{eqnarray}
with
\begin{eqnarray}
&& \chi_{mass}(z)\Big|_{z \ll 1} \simeq G_N s \frac{\pi^{d/2-1}
\Gamma(d/2)}{8} z^{-d} - G_N s \, \frac{\pi^{d/2-1 }}{4 \Gamma(1 +
d/2)}
\nonumber \\
&& \times \left\{ \ln \left( \frac{1}{z} \right) \, \Big[ 1 +
\frac{d}{d + 2} \, z^2 \Big] + \frac{1}{d} + \Psi(1) \right\}
\label{120}
\end{eqnarray}
and
\begin{equation}
\chi_{mass}(z) \Big|_{z \gg 1} = G_N s \frac{\pi^{d/2-1}
\Gamma(d/2)}{4} z^{-d} \sin^2 \Big( \frac{\pi d}{4} \Big).
\label{122}
\end{equation}
We have introduced a dimensionless variable $z = b/2R_c$. The
amplitude $A^{eik}(s,t)$ is well-defined for all $d \geqslant 0$.

Our expression (\ref{118}) has infinite phase (\ref{116}). It was
shown many years ago~\cite{Weinberg:65} that in quantum gravity
each different particle pair in the initial (or final) state
contributes a divergent phase factor to the $S$-matrix.

Note, the second term in (\ref{118}) is regular at $t = 0$. Thus,
for small $t$ (namely, at $|t|R_c^2 \ll 1$) the main contribution
to the eikonal amplitude comes from the Born pole (the first term
in (\ref{118})). It is interesting that this term does not depend
on neither the compactification radius $R_c^2$ or the number of
the extra dimensions $d$. In other words, it is entirely
4-dimensional.

Large $t$ behavior is determined by small values of variable $z$
in integral (\ref{118}). Taking into account the asymtotics of
$\chi_{mass}(z)$ at $z \approx 0$ (\ref{120}), one can obtain by
using stationary-phase technique ($d>0$):
\begin{eqnarray}
&& A^{eik}(s,t) \Big|_{|t| R_c^2 \gg 1}  \simeq -4\pi i s \, e^{i
\phi_4} \, \frac{1}{|t|} \, \frac{ 2^{iG_N s/4 \pi}}{\sqrt{1+d}}
\nonumber \\
&& \times \Big[ \Big( d \big( b_c \sqrt{|t|} \big)^d
\Big)^{1/(d+1)} \Big]^{1 - iG_N s/4 \pi} \exp \Big[ i (1 + d)
\Big( \frac{b_c \sqrt{|t|}}{d} \Big)^{d/(d+1)} \Big].
\label{124}
\end{eqnarray}

Formula (\ref{118}) has correct physical limits. In a case, when
the compactification radius $R_c$ tends to zero and, consequently,
the KK graviton excitations become very heavy ($m_n \rightarrow
\infty$, see (\ref{12})) and decouple from the brane particles,
they make no contribution to the amplitude (since $\chi_{mass}
\rightarrow 0$), but a renormalized Born amplitude still present
in (\ref{118}). The same is true for $d \rightarrow 0$ (no extra
dimensions and, consequently, no massive gravitons are present in
the nature).

On the other hand, the expression for $A^{eik}_{_{GRW}}(s,t)$
obtained in Ref.~\cite{Giudice:02} (see formulae (\ref{44}),
(\ref{48})-(\ref{50})), results in an incorrect value,
$A^{eik}_{_{GRW}}(s,t) = 0$, in the limit $R_c \rightarrow 0$.

\section{Discussions}

In the present paper the scattering amplitude of two \emph{brane
particles}, interacting by the gravity forces, is calculated in
the ADD model in the eikonal approximation. A particular attention
is payed to the account of the contribution from the massless
graviton mode, contrary to a technique used in
Ref.~\cite{Giudice:02}. Our main results is formula (\ref{118}),
where $\chi_{mass}(b)$ represents the contribution from the KK
graviton modes with $n \geqslant 1$. The expression for
$\chi_{mass}(b)$ and its asymptotic behavior are presented by
equations (\ref{88}), (\ref{72}) and (\ref{120}), (\ref{122}),
respectively. Our formula (\ref{118}) gives correct
four-dimensional result at both $D \rightarrow 4$ and $R_c
\rightarrow 0$.

It is interesting to compare our results with those, describing a
collision of two \emph{bulk particles} in $D$ dimensions with
$D>4$. First of all, $\chi^{eik}_{_{ACV}}$ has an imaginary part,
while our $\chi^{eik}$ does not. The imaginary part appears in
$\chi^{eik}$, when one sums multiple exchange of \emph{reggeized
gravitons}~\cite{Muzinich:88,Petrov:02}.

The asymptotics of our eikonal at large impact parameter
(\ref{122}) coincides with the real part of $\chi^{eik}_{_{ACV}}$
(\ref{22}), up to a constant depending on the number of the extra
dimensions.

The $D$-dimensional eikonal amplitude, $A^{eik}_{_{ACV}}$,
(\ref{24}) has the nonrenormalized Born pole at $t=0$. In the
brane amplitude (\ref{118}) the renormalized Born pole is
reproduced. This singular part makes a leading contribution at
small momentum transfers and it coincides with the $D$-dimensional
amplitude taken at $D \rightarrow 4$.

To summarize, the presence of the compact extra dimensions do not
influence small $t$ behavior of the scattering amplitude, if a
collision takes place on the $(1+3)$-dimensional brane. This is
easy to understand, since, from the point of view of four
dimensions, the higher space dimensions supply us with the KK
tower of \emph{massive exchange quanta} (in our case, massive
gravitons). These new massive quanta can not  hide the long-range
forces originated from the massless graviton.

Accounting for the contributions of the massive gravitons results
in an additive term, which is important at large and intermediate
$t$ (\ref{124}). Note, that the eikonal depends, in general, on
the ratio $b/2R_c$ (it is of the form $\chi(b) \sim (b_c/b)^d$
only at $b \rightarrow 0, \, \infty$). Therefore, \emph{the
relevant dimensionless parameters for the amplitude are $G_N s$
and $R_c^2|t|$} (but not $b_c^2|t|$). Remind that in the flat
$(4+d)$ dimensions the eikonal amplitude is given in terms of the
dimensional parameter $G_D s |t|^{d/2}$ (see formulae in Section
\ref{sec:bulk}).

Let us stress again, we disagree with the statement of
Refs.~\cite{Giudice:02,Giudice*:02} saying that the brane
amplitude in the eikonal approximation is dominated by the region
$b \sim b_c$, and it has no Coulomb part (see our comments in the
text after formula (\ref{51})). The accurate account of the
massless graviton mode shows that this is not the case, and
\emph{$A^{eik}$ has both ``massless'' (Coulomb) and ``massive''
term} (\ref{118}).

In Ref.~\cite{Giudice*:02} the quantity $\chi_{_{GRW}}(b)$
from~\cite{Giudice:02} was applied for calculations of di-jet
differential cross sections at the LHC energy in a kinematical
region where gravity dominates. For not small $t$, the
``massless'' part of the eikonal  can be neglected. Nevertheless,
the ``massive'' term of the eikonal, representing the contribution
from the KK graviton modes, is given by more complicated
expression than that derived in~\cite{Giudice:02}. In particular,
asymptotic behavior of $\chi(b)$ (\ref{122}) differs from the
corresponding asymptotics for $\chi_{_{GRW}}$(\ref{48}) by a
factor $2\sin^2(\pi d/4)$. It seems, that the estimates for the
di-jet differential cross sections should be reconsidered.

\section*{Acknowledgments}

The author would like to thank V.A.~Petrov for useful discussions
and valuable remarks. He is also indebted to the High Energy
Section of the ICTP, where this paper was partially done, for a
warm hospitality.


\setcounter{equation}{0}
\renewcommand{\theequation}{A.\arabic{equation}}

\section*{Appendix}

In this Appendix we will calculate the asymptotics of the RHS of
equation (\ref{78}) at large values of $b/2R_c$. The quantity
under consideration, $I(b)$, is represented in terms of the
generalized hypergeometric functions, ${}_1F_2 \left( \alpha_1;
\,\beta_1, \beta_2; \, z \right)$ and ${}_2F_3 \left( \alpha_1,
\alpha_2; \, \beta_1, \beta_2, \beta_3; \, z \right)$ (\ref{74}),
(\ref{78}). Let us introduce the notations
\begin{equation}
{}_pF_q \left( \left.
\begin{array}{c}
  \alpha_p \\
  \beta_q \\
\end{array} \right| z
\right) \equiv {}_pF_q \left( \alpha_1, \ldots \alpha_p;
\,\beta_1, \ldots \beta_q; \, z \right).
\label{A02}
\end{equation}
\begin{equation}
\Gamma(\alpha_p) \equiv \Gamma(\alpha_1) \Gamma(\alpha_2) \ldots
\Gamma(\alpha_p).
\label{A04}
\end{equation}
Sometimes we will write simply ${}_pF_q(z)$.

To solve the problem, we need to use a full asymptotic expansion
of ${}_pF_{p+1}(z)$ at large $z$~\cite{Luke}:
\begin{eqnarray}
&& {}_pF_{p+1} \left( \left.
\begin{array}{c}
  \alpha_p \\
  \beta_{p+1} \\
\end{array} \right| \frac{z^2}{4} \right)
\simeq \frac{\Gamma(\beta_{p+1})}{\Gamma(\alpha_p)} \left[\{
K_{p,p+1} \Big[ \Big( \frac{1}{2}z \Big)^2
\Big] \right.
\nonumber \\
&& + \left. K_{p,p+1} \Big[ \Big(\frac{1}{2}z e^{i\pi} \Big)^2
\Big] + L_{p,p+1} \Big[ \Big( \frac{1}{2}z e^{i\pi} \Big)^2 \Big]
\right\},
\label{A06}
\end{eqnarray}
$|z| \rightarrow \infty, \ \arg z = 0$. Let us remind that $z =
b/R_c > 0$ and $p = 1,\,2$ in our case. The function
$K_{p,p+1}(z)$ in (\ref{A06}) is given by the series in inverse
powers of variable $z$~\cite{Luke}:
\begin{equation}
K_{p,p+1} \Big[ \Big( \frac{1}{2}z \Big)^2 \Big] =
\frac{1}{2^{2\gamma + 1} \sqrt{\pi}} e^z z^{2\gamma}
\sum_{k=0}^{\infty} d_k z^{-k}, \quad d_0 = 1,
\label{A08}
\end{equation}
with
\begin{equation}
\gamma = \frac{1}{2} \left( \frac{1}{2} + \sum_{n=1}^p \alpha_n  -
\sum_{n=1}^{p+1} \beta_n \right).
\label{A10}
\end{equation}
Thus, $K_{p,p+1}(z)$ increases exponentially in $z$ at $z
\rightarrow +\infty$, that can result in an exponential rise of
$I_{\epsilon}$ (\ref{78}) in the impact parameter $b$. However, we
will show now that it is not the case due to \emph{a complete
cancellation} of terms, proportional to $K_{1,2}(z)$ and
$z^{\epsilon} K_{2,3}(z)$ in (\ref{78}).

Let us note first that $\gamma = -3/4$ and $\gamma = -3/4 -
\epsilon $, respectively, for ${}_1F_2 \left( d/2 - \epsilon; 1 +
d/2 - \epsilon, 1;\, z \right)$ and ${}_2F_3 \left( d/2, 1; 1 +
d/2, 1 + \epsilon, 1 + \epsilon;\, z \right)$. For $p = 1$,
coefficients $d_k$ in (\ref{A08}) obey recursion
formulae~\cite{Luke}:
\begin{eqnarray}
&& 2(k + 1)\,d_{k+1}^{(p=1)} = [3k^2 +2k(1 +C_1 -3B_1) + 4D_1]
\,d_k^{(p=1)}
\nonumber \\
&& -(k - 2\gamma - 1)(k - 2\gamma + 1 - 2\beta_1)(k - 2\gamma + 1
- 2\beta_2) \,d_{k-1}^{(p=1)}
\nonumber \\
&& = \tilde{A}_k^{(p=1)} \, d_k^{(p=1)} + \tilde{A}_{k-1}^{(p=1)}
\, d_{k-1}^{(p=1)},
\label{A12}
\end{eqnarray}
where we have introduced the notations
\begin{equation}
B_1 = \sum_{n=1}^p \alpha_n, \qquad C_1 = \sum_{n=1}^{p+1}
\beta_n,
\label{A14}
\end{equation}
\begin{equation}
B_2 = \sum_{n=2}^p \sum_{m=1}^{n-1} \alpha_n \alpha_m, \qquad C_2
= \sum_{n=2}^{p+1} \sum_{m=1}^{n-1} \beta_n \beta_m,
\label{A16}
\end{equation}
\begin{equation}
D_1 = C_2 - B_2 + \frac{1}{4} (B_1 - C_1)(3B_1 + C_1 -2) -
\frac{3}{16}.
\label{A17}
\end{equation}

For $p = 2$, recursion relations look like~\cite{Luke}:
\begin{eqnarray}
&& 2(k + 1)\,d_{k+1}^{(p=2)} = [5k^2 + 2k(3 + B_1 - 3C_1
-10\gamma) +4D_1] \,d_k^{(p=2)} \nonumber \\
&& - [4k^3 - 6k^2(C_1 + 4\gamma) + 2k(24\gamma^2 + 12\gamma C_1 +
C_1 +4C_2 - 1)
\nonumber \\
&& - 32\gamma^3 - 24\gamma^2C_1 -4\gamma(C_1 + 4C_2 - 1) + 2C_1 -
4C_2 - 8C_3 -1] \, d_{k-1}^{(p=2)}  \nonumber \\
&& + (k - 2\gamma - 2)(k - 2\gamma - 2\beta_1)(k - 2\gamma -
2\beta_2)(k - 2\gamma - 2\beta_3)\,d_{k-2}^{(p=2)}
\nonumber \\
&& = A_k^{(p=2)} \, d_k^{(p=2)} + A_{k-1}^{(p=2)}
\,d_{k-1}^{(p=2)} +  A_{k-2}^{(p=2)} \,d_{k-2}^{(p=2)},
\label{A18}
\end{eqnarray}
where
\begin{equation}
C_3 = \beta_1 \beta_2 \beta_3
\label{A20}
\end{equation}
and other quantities are defined as before
(\ref{A14})-(\ref{A17}).

By making a replacement $k \rightarrow k-1$ in (\ref{A12}), we get
\begin{eqnarray}
&& 2k\,d_k^{(p=1)} = [3(k - 1)^2 +2(k - 1)(1 +C_1 -3B_1) + 4D_1]
\,d_{k-1}^{(p=1)}
\nonumber \\
&& -(k - 2\gamma - 2)(k - 2\gamma  - 2\beta_1)(k - 2\gamma -
2\beta_2) \,d_{k-2}^{(p=1)}.
\label{A22}
\end{eqnarray}
Let us now rewrite equation (\ref{A12}) in the form
\begin{eqnarray}
2(k + 1)\,d_{k+1}^{(p=1)} &=& [\tilde{A}_k^{(p=1)} - A_k^{(p=2)}]
\, d_k^{(p=1)}
\nonumber \\
&+& A_k^{(p=2)} \, d_k^{(p=1)} + \tilde{A}_{k-1}^{(p=1)} \,
d_{k-1}^{(p=1)}
\label{A24}
\end{eqnarray}
and substitute $d_k^{(p=1)}$ from (\ref{A22}) into the first term
in the RHS of (\ref{A24}). Then we obtain:
\begin{equation}
2(k + 1)\,d_{k+1}^{(p=1)} = A_k^{(p=1)} \, d_k^{(p=1)} + A_{k-1}^{(p=1)}
\,d_{k-1}^{(p=1)} +  A_{k-2}^{(p=1)} \,d_{k-2}^{(p=1)}.
\label{A26}
\end{equation}
By direct calculations one can check that $A_i^{(p=1)} =
A_i^{(p=2)}$, namely:
\begin{eqnarray}
A_k^{(p=1)} = A_k^{(p=2)} &=& 5k^2 + k(5 - 2n + 8\epsilon)
+\frac{13}{4} -2n +4\epsilon,
\nonumber \\
A_{k-1}^{(p=1)} =  A_{k-1}^{(p=2)} &=& -4k^3 +3k^2(n - 4\epsilon)
+ k(-1 + 4n\epsilon -8\epsilon^2)
\nonumber \\
&+& \frac{1}{4}(n - 4\epsilon),
\nonumber \\
A_{k-2}^{(p=1)} =  A_{k-2}^{(p=2)} &=&  \frac{1}{16}(2k - 1)^2(2k
-1 + 4\epsilon) \nonumber \\
&\times& (2k -1 -2n + 4\epsilon).
\label{A28}
\end{eqnarray}
In other words, we have shown that both $d_k^{(p=1)}$ and
$d_k^{(p=2)}$ obeys the same recursion formulae, and $d_k^{(p=1)}
= d_k^{(p=2)}$ \emph{for all $k$} in series (\ref{A08}).

Therefore, the $K$-functions do not contribute to expansion
(\ref{A06}) in our case, and we have
\begin{eqnarray}
I_{\epsilon}(b) &=& \frac{d}{2\epsilon} \, \Gamma(1 + \epsilon)
\Gamma(1 + \epsilon) \left[ - \left( \frac{b^2}{4 R_c^2}
\right)^{\epsilon}
\right. \nonumber \\
&\times& L_{2,3} \left( \left.
\begin{array}{c}
  d/2,\,1 \\
  1+d/2,\,1+\epsilon,\,1+\epsilon \\
\end{array}
\right| \frac{b^2}{4 R_c^2} \right) \nonumber \\
&+& \left.
 L_{1,2} \left( \left.
\begin{array}{c}
  d/2-\epsilon \\
  1+d/2-\epsilon,/,1 \\
\end{array}
\right| \frac{b^2}{4 R_c^2} \right) \right].
\label{A30}
\end{eqnarray}
Note, the $L$-function is related to the Meijer
$G$-function~\cite{Luke}:
\begin{equation}
L_{p,p+1} \left( \left.
\begin{array}{c}
  \alpha_1, \ldots \alpha_p \\
  \beta_1, \ldots \beta_{p+1} \\
\end{array}
\right| z \right) = G_{p+2,\,p}^{p,\,1} \left( \frac{1}{z} \left|
\begin{array}{c}
  1,\,\beta_1, \ldots \beta_{p+1} \\
  \alpha_1, \ldots \alpha_p \\
\end{array}
\right. \right).
\label{A32}
\end{equation}
In its turn, the $G$-function can be represented as a series in
generalized hypergeometric functions of an inverse power of
$z$~\cite{Bateman}:
\begin{equation}
L_{p,p+1}(z) = \sum_{n=1}^p L_{p,p+1}^{(n)}(z),
\label{A34}
\end{equation}
where
\begin{eqnarray}
L_{p,p+1}^{(n)}(z) &=& z^{-\alpha_n} \frac{\Gamma(\alpha_n)
\Gamma( \alpha_p - \alpha_n)^*}{\Gamma( \beta_{p+1} - \alpha_n)}
\nonumber \\
&\times&_{p+2}F_{p-1} \left( \left.
\begin{array}{c}
  \alpha_n,\, 1 + \alpha_n - \beta_{p+1} \\
   1 + \alpha_n - \alpha_p^*
\end{array} \right| - \frac{1}{z}
\right)
\label{A36}
\end{eqnarray}
(the superscript ${}^*$ means that the term with $\alpha_n =
\alpha_p$ is not included in a product of the $\Gamma$-functions).

As a resits, we arrive at the expression for $I_{\epsilon}$, which
is convenient for analyzing its large $b$ behavior:
\begin{eqnarray}
I_{\epsilon}(b) &=& \frac{1}{\epsilon} \, \Gamma(1 - \epsilon)
\Gamma(1 + \epsilon) \Gamma(1 + d/2)
\nonumber \\
&\times& \left\{ - \left( \frac{b^2}{4 R_c^2} \right)^{\epsilon -
1} \frac{ \Gamma(d/2 - 1)}{\Gamma^2(\epsilon) \Gamma^2(d/2)}
\right.
\nonumber \\
&\times& {}_4F_1 \left(1, 1 - \frac{d}{2}, 1 - \epsilon,   1 -
\epsilon; 2 - \frac{d}{2}; - \frac{4R_c^2}{b^2} \right)
\nonumber \\
&+& \left( \frac{b^2}{4 R_c^2} \right)^{\epsilon - d/2}
\frac{1}{\Gamma(1 - d/2 +\epsilon)}
\nonumber \\
&\times& \left. \left[ - \frac{\Gamma(1 - d/2)}{\Gamma(1 - d/2 +
\epsilon)}
 +  \frac{\Gamma(d/2 - \epsilon)}{\Gamma(d/2)} \right] \right\}.
\label{A38}
\end{eqnarray}
Up to now, we did not consider parameter $\epsilon$ to be small.
Finally, a desired asymptotics looks like
\begin{eqnarray}
I(b) \Big|_{b \gg R_c} &=& \lim_{\epsilon \rightarrow 0}
I_{\epsilon}\Big|_{b \gg R_c} = \left( \frac{2 R_c}{b} \right)^d
\, \frac{\Gamma(1 + d/2)}{\Gamma(1 - d/2)}
\nonumber \\
&\times& \, \lim_{\epsilon \rightarrow 0} \frac{1}{\epsilon}
\left[ - \frac{\Gamma(1 - d/2)}{\Gamma(1 - d/2 + \epsilon)} +
\frac{\Gamma(d/2 - \epsilon)}{\Gamma(d/2)} \right].
\label{A40}
\end{eqnarray}
By expanding the RHS of equality (\ref{A40}) in $\epsilon$ and
taking the limit $\epsilon \rightarrow 0$, we derive an asymptotic
formula for $I(b)$ presented in the text (\ref{82}).

\end{document}